\documentclass[twoside]{article}

\usepackage[sc]{mathpazo} 
\usepackage[T1]{fontenc} 
\usepackage{microtype} 

\usepackage[hmarginratio=1:1,top=32mm,columnsep=20pt]{geometry} 
\usepackage{multicol} 
\usepackage[hang, small,labelfont=bf,up,textfont=it,up]{caption} 
\usepackage{booktabs} 
\usepackage{float} 
\usepackage{hyperref} 

\usepackage{lettrine} 
\usepackage{paralist} 

\usepackage{abstract} 

\usepackage{titlesec} 
\renewcommand\thesection{\Roman{section}}
\titleformat{\section}[block]{\large\scshape\centering}{\thesection.}{1em}{} 

\usepackage{fancyhdr} 
\usepackage{amssymb}
\usepackage{epsfig}
\usepackage{amssymb}
\usepackage{url}
\usepackage{amsmath}
\usepackage{natbib}
\usepackage{amsthm}
\usepackage{booktabs}
\usepackage[lined,boxed,linesnumbered]{algorithm2e}
\usepackage{url}
\usepackage{alltt}
\usepackage{enumerate}
\usepackage{multirow}
\usepackage{setspace}
\usepackage{bm}
\usepackage{color}

\newcommand{\bbE}{\mathbb{E}}

\newcommand{\bbR}{\mathbb{R}}

\newcommand{\sX}{\mathcal{X}}
\newcommand{\sY}{\mathcal{Y}}
\newcommand{\sZ}{\mathcal{Z}}
\newcommand{\ry}{\mathrm{y}}
\newcommand{\vx}{\mathbf{x}}
\newcommand{\vy}{\mathbf{y}}
\newcommand{\PD}{\psi}   
\newcommand{\vz}{\mathbf{z}}

\newcommand{\pr}{\mathtt{Prob}}

\newcommand{\Real}{I \! \! R}

\def\boldfacefake#1{\kern-4pt
    \hbox{ \mathsurround=0pt
    \hbox to 0.4pt{$#1$\hss}\hbox to 0.4pt{$#1$\hss}\hbox {$#1$}}}

\newcommand{\be}{\begin{eqnarray}}
\newcommand{\ee}{\end{eqnarray}}
\newcommand{\ba}{\begin{eqnarray*}}
\newcommand{\ea}{\end{eqnarray*}}

\newcommand{\mX}{\mathbf{X}}
\newcommand{\mY}{\mathbf{Y}}
\newcommand{\mI}{\mathbf{I}}
\newcommand{\mA}{\mathbf{A}}
\newcommand{\mB}{\mathbf{B}}

\newtheorem{theorem0}{Theorem}
\newtheorem{lemma0}{Lemma}
\newtheorem{remark0}{Remark}
\newtheorem{fact0}{Fact}
\newtheorem{example0}{Example}
\newtheorem{definition0}{Definition}
\newtheorem{corollary0}{Corollary}
\newtheorem{proposition0}{Proposition}
\newtheorem{algorithmY}{Algorithm}
\newtheorem{conjecture0}{Conjecture}

\newenvironment{theorem}{\begin{theorem0} \mbox{} }{\end{theorem0}}

\newenvironment{definition}{\begin{definition0}
\mbox{}}{\end{definition0}}

\newenvironment{proposition}{\begin{proposition0}\mbox{}
}{\end{proposition0}}

\title{\vspace{-15mm}\fontsize{19pt}{10pt}
\selectfont\textbf{To Bayes or Not To Bayes?\\ {\it That's no longer the question.}}} %

\author{
\textbf{Ernest Fokou\'e}\\
\normalsize School of Mathematical Sciences \\ 
\normalsize Rochester Institute of Technology \\ 
\normalsize 98 Lomb Memorial Drive \\
\normalsize Rochester, NY 14623, USA \\ 
\normalsize \href{mailto:epfeqa@rit.edu}{epfeqa@rit.edu} 
\vspace{-5mm}
}
\date{}

\begin{document}

\maketitle 

\thispagestyle{fancy} 

\begin{abstract}
This paper seeks to provide a thorough account of the ubiquitous nature of the Bayesian paradigm in modern statistics, data science and artificial intelligence. Once maligned, on the one hand by those who philosophically hated the very idea of subjective probability used in prior specification, and on the other hand because of the intractability of the computations needed for Bayesian estimation and inference, the Bayesian school of thought now permeates and pervades virtually all areas of science, applied science, engineering, social science and even liberal arts, often in unsuspected ways. Thanks in part to the availability of powerful computing resources, but also to the literally unavoidable inherent presence of the quintessential building blocks of the Bayesian paradigm in all walks of life, the Bayesian way of handling statistical learning, estimation and inference is not only mainstream but also becoming the most central approach to learning from the data. This paper explores some of the most relevant elements to help to the reader appreciate the pervading power and presence of the Bayesian paradigm in statistics, artificial intelligence and data science, with an emphasis on how the Gospel according to Reverend Thomas Bayes has turned out to be the truly good news, and some cases the amazing saving grace, for all who seek to learn statistically from the data. To further help the reader gain deeper and tangible practical insights into the Bayesian machinery, we point to some computational tools designed for the R Statistical Software Environment to help explore Bayesian statistical learning.
\end{abstract}

\section{Introduction}
{\it "Dear Sir, I now send you an essay which I have found among the papers of our deceased friend Mr. Bayes, and which, in my opinion, has great merit, and well deserves to be preserved."} This first sentence of the cover letter written by Mr John Canton F.R.S., who honorably introduced the manuscript of the phenomenal work of the by then Late Reverend Thomas Bayes F.R.S, that first sentence said I, could not have been more accurate when it said "{\it has great merit, and well deserves to be preserved}". $254$ years later, it is understatement to say that the work of Reverend Thomas Bayes, introduced and developed in that manuscript, does permeate virtually every aspect of scientific analysis involving the doctrine of chance and probability. It is so rich indeed in positively transformative concepts that it won't be an exaggeration to refer to it as the {\it gospel according to Reverend Thomas Bayes}, judging by the sheer plurality of its applications to literally all areas of statistical and probabilistic modelling. As a matter of fact, both explicitly and implicitly, an overwhelmingly large number of the so-called learning machines in artificial intelligence, statistical machine learning or data science, admit a Bayesian formulation often directly or after simple transformations. The multiplicity of such occurrences leads one to recognize the quasi-centrality of the Bayesian paradigm in science in general. Indeed, {\it To Bayes or Not To Bayes?} is no longer the question, but rather {"How to Bayes?}, since the Bayesian paradigm appears ubiquitous, permeating and pervading every scientific activity involving the doctrine of chance and statistical learning from the data. In an era marked by the resurgence of artificial intelligence\footnote{{\tt Both the well known Weak Artificial Intelligence, and the highly anticipated Strong AI.}} and the firm establishment of statistical machine learning as a force to reckon with, along with the meteoric rise to prominence of the emerging field of data science, all of which have to deal with uncertainty at their core, it makes sense the statistics, the natural language (along with sister probability) for dealing with uncertainty, should permeate the very fabric of epistemology, theory, methodology, computation and application. Interestingly, as we will see later, the famous Bayes' theorem (Bayes' rule or Bayes' formula as some call it) stands prominently and firmly at the very core, providing a versatile, rich and powerful paradigm for modelling both the simplest and the most complex of phenomena. From the fundamental algebra of finite sets of events to the estimation of model parameters to infinite dimensional function approximation and estimation, the Bayesian paradigm seems to find a way to emerge (sometimes almost miraculously) as the de-facto flexible modelling framework for formulating and/or solving the task at hand. The goal of this paper is not to preach the Gospel according Reverend Thomas Bayes, not is it aimed at reviewing the sophisticated technical niceties of some seminal Bayesian fundamental results. Instead, our goal is to provide a general bird's eye view of the manifold incarnations of the Bayesian machinery in artificial intelligence, statistical machine learning and data science. In its most generic and canonical form, Bayes' theorem is used to connect the conditional and marginal probabilities of two events.
\begin{theorem}
Let $\mA$ and $\mB$ be two events with nonzero probabilities such $\Pr(\mA) > 0$ and $\Pr(\mB)$, then the conditional probability of $\mB$ given that $\mA$ has occurred, is given by
\begin{eqnarray}
\label{thm:Bayes:basic:1}
\Pr(\mB|\mA) = \frac{\Pr(\mB)\Pr(\mA|\mB)}{\Pr(\mA)}=\frac{\Pr(\mB)\Pr(\mA|\mB)}{\Pr(\mB)\Pr(\mA|\mB)+\Pr(\mB^c)\Pr(\mA|\mB^c)}.
\end{eqnarray}
\end{theorem}

An extension deals with a collection $\mB_1, \cdots, \mB_K \in \Omega$ be mutually exclusive events, and their probabilistic relationship with some event $\mA \in \Omega$.

\begin{theorem}
Let $\mA \in \Omega$ be an event with nonzero probability such $\Pr(\mA) > 0$, and  consider the collection of mutually exclusive events $\mB_1, \cdots, \mB_K \in \Omega$ such that  $\mB_k \cap \mB_j = \emptyset,\, j\neq k$ and $\sum_{k=1}^K{\Pr(\mB_k)}=1$, ie $\cup \mB_k = \Omega$, then the conditional probability of $\mB_k$ given that $\mA$ has occurred, is given by
\begin{eqnarray}
\label{thm:Bayes:basic:2}
\Pr(\mB_j|\mA) = \frac{\Pr(\mB_j)\Pr(\mA|\mB_j)}{\Pr(\mA)}=\frac{\Pr(\mB_j)\Pr(\mA|\mB_j)}{\sum_{k=1}^K{\Pr(\mB_k)\Pr(\mA|\mB_k)}}.
\end{eqnarray}
\end{theorem}

The central tenant of the Bayesian paradigm is the concept of posterior probability of an event.
For instance, $\Pr(\mB_j|\mA)$ in \eqref{thm:Bayes:basic:2} is the posterior probability of event
$\mB_j$ given $\mA$, which measures the probability that $\mB_j$ will occur, given that $\mA$ has occurred.
This concept of posterior probability provides a powerful mechanism for formulating, modelling and computing
prediction and predictive quantities of all kinds. It is important however to emphasize that prediction here is not forecasting, nor is it meant in the sense of causation. Prediction here is meant in the sense of dependent arising.
In Bayesian parlance, $\Pr(\mB_j)$ represents the prior belief in $\mB_j$ before the dependent event $\mA$ occurs, and in that sense, the posterior $\Pr(\mB_j|\mA)$ updates the belief in $\mB_j$ given that $\mA$ has occurred.
$\Pr(\mA)$ is referred to as the evidence by many in the Bayesian community, enjoys that appellation most appropriately
in settings like Bayesian hypothesis testing where $\Pr({\bf H}_0|\, {\tt data})$ measures the probability that the null hypothesis is true given the evidence provided by the data. Indeed this concept of evidence is key for a variety of reasons.

\section{Bayes' Impact in Statistical Learning Theory}
To help clarify all the above claims, let $\sX$ and $\sY$ be two sets, and consider their Cartesian product $\sZ \equiv \sX \times \sY$. Now, define $\sZ^n \equiv \sZ \times \sZ \times \cdots \times \sZ$ to be the $n$-fold cartesian product of $\sZ$. Assume that $\sZ$ is equipped with a probability measure $\PD$, and let $\vz \in \mathcal{Z}^n, \,\, \text{with}\,\, \vz  = ((\vx_1,y_1),(\vx_2,y_2)\cdots,(\vx_n,y_n))$
denote the realization of a random sample of $n$ examples, where each example $z_i=(\vx_i,y_i)$  is independently drawn
according to the above probability measure $\PD$ on the product space $\mathcal{Z} \equiv \mathcal{X} \times \mathcal{Y}$. Now, {\it given a random sample
$\vz=((\vx_1,y_1), (\vx_2,y_2), \cdots,(\vx_n,y_n))$ and assuming that the probability measure $\PD$ is unknown,
find the function $f:\sX \rightarrow \sY$ that best captures the dependencies between the
$\vx_i$'s and the $y_i$'s.} We shall refer to $\sX$ as the input space, and to $\sY$ as the output space.
For simplicity, we shall assume  that $\sX \subseteq \bbR^p$, and we shall also consider both regression corresponding to $\sY = \Real$, and classification (pattern recognition) corresponding to $\sY = \{c_1, c_2, \cdots, c_g,\cdots, c_G\}$. This setting where one seeks to estimate a function $f: \sX \rightarrow \sY$, is the foundational setting of machine learning in general and statistical machine learning in particular. This setting brings with the need to extend Bayes' theorem from events to random variables, especially with concepts of marginal density, conditional density and conditional expectation. We see here that the Bayesian paradigm provides the perfect mechanism for the most fundamental
results in pattern recognition, regression, hypothesis testing, signal detection, parameter estimation, function estimation  and statistical learning in general.
In binary classification (pattern recognition), the Bayesian framework provides a convenient language in the assessment of a classifier $f: \sX \rightarrow \sY$, namely for computing the so-called True Positive Rate ({\tt TPR}) and False Positive Rate ({\tt FPR}) which both use the concept of posterior probability. Indeed, the True Positive Rate ({\tt TPR}) of $f$ is given by
$$
{\tt TPR}(f) = \Pr(f(X)=1|Y=1) = \frac{\Pr(f(X)=1 \, \text{and} \, Y=1)}{\Pr(Y=1)},
$$
and  the False Positive Rate ({\tt FPR}) of $f$ is given by
$$
{\tt FPR}(f) = \Pr(f(X)=1|Y=-1) = \frac{\Pr(f(X)=1 \, \text{and} \, Y=-1)}{\Pr(Y=-1)}.
$$
In Bayesian hypothesis testing, the decision about the null hypothesis is conveniently made by measuring the posterior probability of ${\tt H}_0$ given the data, which is given by
$$
\Pr({\bf H}_0|\, Y=\vy) = \frac{p_Y(\vy |{\bf H}_0) \Pr({\bf H}_0)}{p_Y(\vy)}
$$
where $\Pr({\bf H}_0) + \Pr({\bf H}_a) =1$ and  $p_Y(\vy) = \Pr({\bf H}_0)p_Y(\vy |{\bf H}_0)
+ p_Y(\vy |{\bf H}_a)\Pr({\bf H}_a)$ is the density of the data. It is interesting to note the error of the test is also conveniently defined and calculated as
$$
{\tt Error} = \Pr({\tt H}_0\, {\tt \, is \, chosen} | {\tt H}_a)\Pr({\tt H}_a)
+ \Pr({\tt H}_a\, {\tt \, is \, chosen} | {\tt H}_0)\Pr({\tt H}_0).
$$

When we consider the pattern recognition task once again along with the so-called $0/1$ loss function defined below, we encounter another fundamental pattern recognition result that has its foundation in the Bayesian paradigm.   More specifically, consider the $0/1$ loss function
\begin{equation}
\label{eq:01loss}
\ell(y,f(\vx)) = {\bf I}(y \neq f(\vx)) = \left\{
                       \begin{array}{ll}
                         0 & \hbox{if} \quad y = f(\vx),\\
                         1 & \hbox{if} \quad y \neq f(\vx).
                       \end{array}
                     \right.
\end{equation}
With the zero-one loss function in classification, our corresponding {\sf true risk} (also known
as {\sf theoretical risk} or {\sf generalization error} or {\sf true error}) is given by
\begin{equation}
\label{eq:true:risk:1}
R(f) = \int{\ell(y,f(\vx)) d \PD(\vx,y)} = \mathbb{E}\left[{\bf I}(Y\neq f(X))\right] =
\pr_{(X,Y)\sim \PD}[Y\neq f(X)].
\end{equation}
The true error $R(f)$ of a classifier $f$ therefore defines the probability that $f$ misclassifies any arbitrary observation randomly draw from
the population of interest according to the distribution $\PD$. It is important to note from the definition that
$R(f)$ can also be interpreted as the {\it expected} disagreement between classifier $f$ and
the truth about the label $y$ of $\vx$.

\begin{definition} {\bf How is the Bayes' classifier obtained? } Consider a pattern $\vx$ from the input space,  and a class label $y$. Let $p(\vx|y)$ denote the class conditional density of $\vx$ in class $y$, and let $\pr[Y=y]$ denote the prior probability of class membership. The posterior probability of class membership is defined as
$$
\pr[Y = y | \vx] = \frac{\pr[Y=y]p(\vx|y)}{p(\vx)}.
$$
Given a pattern $\vx$ to be classified, the Bayes classification strategy consists of
{\sf assigning  \, $\vx$ \, to the class with \,  maximum \, posterior probability.}
More formally, with $h$ denoting the function from $\mathcal{X}$ to $\sY$ that implements
the Bayes classifier, we have, $\forall \vx \in \sX$,
\begin{eqnarray}
\label{eq:Bayes:classifier}
h(\vx)  &=& \underset{c\in\sY}{\tt argmax}\left\{\pr(Y=c|\vx)\right\}.
\end{eqnarray}
\end{definition}
\begin{theorem}
The minimizer of the $0/1$ risk functional over all possible classifiers is the { Bayes classifier}  $h$ defined in \eqref{eq:Bayes:classifier}.
\begin{eqnarray}
\label{eq:Bayes:classifier:optimal:1}
f^*= \underset{f}{\tt arginf}\left\{R(f)\right\} = \underset{f}{\tt arginf}\left\{\mathbb{E}\left[{\bf I}(Y\neq f(X))\right]\right\} = \underset{f}{\tt arginf}\left\{\pr_{(X,Y)\sim \PD}[Y\neq f(X)]\right\}=h.
\end{eqnarray}
Therefore, the Bayes' classifier $h$ defined in \eqref{eq:Bayes:classifier}, is the universal best classifier, such that $\forall \vx \in \sX$,
\begin{eqnarray}
\label{eq:Bayes:classifier:optimal:2}
f^*(\vx)= h(\vx) = \underset{c\in\sY}{\tt argmax}\left\{\pr(Y=c|\vx)\right\}.
\end{eqnarray}
\end{theorem}
The risk $R^*$ corresponding to $f^*$ is the smallest possible error that any classifier can achieve, i.e.,
$$
R^* = R(f^*) =  R(h) = \underset{f}{\tt inf}\left\{R(f)\right\}.
$$
This result, namely that the {\it Bayes classifier} achieves the universal (global) minimum (infimum) error over all measurable classifiers,  is fundamental result in pattern recognition and statistical learning. The probability theory for pattern recognition is made up of multiple results featuring learning machines whose performance are compared to the performance of the Bayes' classifier. \cite{devroye1997probabilistic} and \cite{Vapnik:00:1}. A similar fundamental statistical learning result exists for regression, namely that under the so-called squared error loss, the universal best function is the conditional expectation of $Y$ given $X$.
\begin{theorem}
Consider functions $f: \Real^p \rightarrow \Real$, and the squared risk functional
$$
R(f) = \mathbb{E}[(Y-f(X)^2] = \int_{\sX \times \sY}{(\ry-f(\vx))^2 p(\vx,\ry) d\vx d\ry}.
$$
Then the best function $f^* = \underset{f}{\tt arginf}\left\{R(f)\right\}$ is given by the conditional expectation of $Y$ given $X$, so that $\forall \vx \in \sX$,
\begin{eqnarray}
\label{thm:Bayes:regressor:1}
f^*(\vx) = \underset{f}{\tt arginf}\left\{R(f(\vx))\right\}=\bbE[Y|X =\vx] = \int_{\mathcal{Y}}{\ry \,p(\ry|\vx) d\ry}
\end{eqnarray}
\end{theorem}
\begin{proposition}
For every $f: \mathcal{X} \rightarrow \mathcal{Y}$,
$$
R(f) = \int_{\mathcal{X}}{(f(\vx)-f^*(\vx))^2 d\, \PD(\vx)} + \sigma_{*}^2.
$$
\end{proposition}
We see that for both regression and classification, the Bayesian paradigm provides the best mechanism, at the very least in theory, which is indeed very important. It is worth mentioning that for most people, the Bayesian school of thought is typically not introduced through results like the ones we just described, but instead through Bayesian estimation and inference in parametric families of models. It is our view that both the Bayes' classifier and the Bayes regressor are just as valuable members of the Bayes' heritage as are the vastly studied results in both parametric and nonparametric Bayesian estimation, inference prediction. Although the results described earlier were in their purely theoretical forms, applications abound that are based on those foundational results. Studying pattern recognition and regression with solid knowledge of both the Bayes' classifier and the Bayes regressor which provide the best in both cases is of vital importance\footnote{Most people think of the Bayesian paradigm the {\sf sensu stricto } where there is a very involved and often complex and  sometimes controversial topic of prior specification. Our treatment of the Bayesian idea is definitely in {\sf sensu lato}, and encompasses all modeling situations where the posterior density or the posterior probability is part of the modeling mechanism. Our intention is not to trigger an epistemological debate, quite far from it.}. It is my view that in {\it sensu lato}, all statistical learning methods are offshoots of the Bayesian machinery in the sense  the Bayesian learner under the two most commonly loss functions is always the optimal, indeed the standard. In that sense, most so-called non-Bayesians or anti-Bayesians are inherently Bayesians at their core, at least in the most quintessential sense of those universal optimality results that all learning machines essentially attempt to attain.

\section{Bayes' Impact in Statistical Estimation and Inference}
Now, the best known setup where the richness of the Bayesian paradigm is practically and more directly revealed, is encountered when we assume that the task of learning the function $f$, is associated with the estimation of a parameter $\theta \in \Theta \subseteq \bbR^p$ such that, when treated as a random variable, the probability density function of $\theta$ is given by $p(\theta)$. This is encountered for models involving (a) Parametric density estimation along with elements of prediction; (b) Parametric function estimation along with prediction, when $f(\vx)=f(\vx; \theta)$. In both cases, a key quantity is the posterior density of the parameter $\theta$ given the data, namely
\begin{eqnarray}
\label{eq:post:theta:1}
p(\theta|\mY) = \frac{p(\theta)p(\mY|\theta)}{p(\mY)},
\end{eqnarray}
where $p(\mY|\theta)$ is the likelihood of $\theta$, and $p(\mY)$ is the evidence, sometimes referred to as the marginal likelihood of the underlying model. Recall that the likelihood of $\theta$ is the joint density of the data vector $\mY$ given the unknown parameter $\theta$, i.e. ${\tt Likelihood}(\theta) = p(\mY|\theta)$. The maximum likelihood principle is arguably the most commonly tool/approach in statistical analysis because of the central role the likelihood plays in statistical modelling. Now, the maximum likelihood estimator $\widehat{\theta}^{(\tt MLE)}$
of $\theta$ is given by
\begin{eqnarray}
\label{eq:mle:theta:1}
\widehat{\theta}^{(\tt MLE)} = \underset{\theta \in \Theta}{\tt argmax}\left\{{\tt Likelihood}(\theta)\right\}= \underset{\theta \in \Theta}{\tt argmax}\left\{p(\mY|\theta)\right\},
\end{eqnarray}
while the Bayesian estimator $\widehat{\theta}^{(\tt Bayes)}$ of $\theta$ under the squared error loss, is
\begin{eqnarray}
\label{eq:bayes:theta:1}
\widehat{\theta}^{(\tt Bayes)} &=& \underset{a \in \Theta}{\tt argmin}\left\{\mathbb{E}[(\theta-a)^2|\mY]\right\} \nonumber \\
&=& \mathbb{E}[\theta|\mY] = \int_{\Theta}{\theta p(\theta|\mY)d\theta}.
\end{eqnarray}
A very nice property of both Maximum Likelihood and Bayesian Estimators is that for all continuous functions $g(\cdot)$,
$$
\widehat{g(\theta)}^{(\tt Bayes)}
= \mathbb{E}[g(\theta)|\mY] = \int_{\Theta}{g(\theta) p(\theta|\mY)d\theta}.
$$
Also the Bayesian paradigm inherently addresses the important predictive density of any new element of $\ry^{(\tt new)} \in \sY$, which is given by
$$
p(\ry_{\tt new} | \mY) = \int_{\Theta}{p(\ry_{\tt new}|\theta)p(\theta|\mY)d\theta}.
$$
{\tt The Maximum A Posteriori (MAP) Estimator} is another type, albeit sometimes deemed inferior, of Bayesian estimator, given by
\begin{eqnarray*}
\widehat{\theta}^{(\tt MAP)} &=& \underset{\theta \in \Theta}{\tt argmax}\left\{p(\theta|\mY)\right\} =
\underset{\theta \in \Theta}{\tt argmax}\left\{p(\theta)p(\mY|\theta)\right\} \\
&=& \underset{\theta \in \Theta}{\tt argmax}\left\{ \log p(\theta)+ \log p(\mY|\theta)\right\}  \\
&=& \underset{\theta \in \Theta}{\tt argmax}\left\{ \log {\tt Prior}(\theta)+ \log {\tt Likelihood}(\theta)\right\}.
\end{eqnarray*}
Note that if the prior density $p(\theta)$ is uniform, i.e. $p(\theta)=c$, then we have
$$
\widehat{\theta}^{(\tt MAP)}= \widehat{\theta}^{(\tt Bayes)} =
\underset{\theta \in \Theta}{\tt argmax}\left\{\log p(\mY|\theta)\right\} =
\underset{\theta \in \Theta}{\tt argmax}\left\{{\tt Likelihood}(\theta)\right\} =
\widehat{\theta}^{(\tt MLE)}
$$
{\it The Bayesian paradigm is therefore an extension and a generalization of the maximum likelihood principle,
an extension that affords greater modelling flexibility, and consequently the capability to solve a wider class of problems. The maximum likelihood estimator is a special case of the Bayesian estimator}.
Another powerful property inherent in the Bayesian paradigm is its inherent shrinkage  and regularization capability, which turns out to be a powerful remedy that helps circumvent a wide variety of modelling challenges.
To gain deeper insights into this regularization and shrinkage property, we consider the Bernoulli experiment, with the parameter $\theta \in (0,1)$ representing the probability of success, and $Y_i \in \{0,1\}$ such that
$$
Y_1, Y_2, \cdots, Y_n \overset{{\tiny iid}}{\sim} {\tt Bernoulli}(\theta)
$$
We have
$$
p(y_i|\theta) = \theta^{y_i} (1-\theta)^{1-y_i}
$$
Under the conjugacy principle, the conjugate prior for $\theta$  is
$$
p(\theta|a,b) = \frac{\Gamma(a+b)}{\Gamma(a)\Gamma(b)}\theta^{a-1}(1-\theta)^{b-1}
$$
It can be shown that the posterior density of $\theta$ is given by
$$
p(\theta|D_n) = \frac{\Gamma(a+b+n)}{\Gamma(S_n+a)\Gamma(n-S_n+b)}\theta^{S_n+a-1}(1-\theta)^{n-S_n+b-1}
$$
Which means that $(\theta|D_n) \sim {\tt Beta}(a+S_n, b+F_n)$. Now we
$$
\widehat{\theta}^{(\tt Bayes)} = \mathbb{E}[\theta|\mY] = \int_{\Theta}{\theta p(\theta|\mY)d\theta}=\frac{a+S_n}{a+b+n}
$$
Notice
$$
\widehat{\theta}^{(\tt Bayes)} =\frac{a+S_n}{a+b+n} = \frac{a+b}{a+b+n}\frac{a}{a+b} + \frac{n}{a+b+n}\frac{S_n/n}{a+b+n}=w_n \widehat{\theta}_0
+ (1-w_n) \widehat{\theta}_{\tiny (\tt MLE)}
$$
The Bayesian "point" estimator $\widehat{\theta}^{(\tt Bayes)}$ is therefore a convex combination of the prior estimate with the maximum likelihood estimator. Indeed, $\underset{n \rightarrow \infty}{\lim} w_n = 0$. As  a result,
$$
\underset{n \rightarrow \infty}{\lim} \widehat{\theta}^{(\tt Bayes)} = \widehat{\theta}^{(\tt MLE)}
$$
Which means that as more data becomes available, the posterior density is dominated by the likelihood, so that the asymptotically the Bayesian estimator coincide with the maximum likelihood estimator. In this sense, the prior is bringing to the estimation (learning) task, items of information that the data from the sampling process does not contain, and this is crucial. As more data becomes available, the information brought by the prior is then overwhelmed by the information richly provided by large amounts of data. This serves as the basis for resorting to the Bayesian paradigm in situations where there isn't enough data to carry the modelling task at hand.

\section{Bayes' Impact in Statistical Function Estimation}
To better understand this, we consider multiple linear regression under the Gaussian homoscedastic noise model, $(\mY|\mX,\theta,\sigma^2) \sim {\bf N}_n(\mX\theta, \sigma^2 \mI_n)$, for which the likelihood of $\theta$ is simply
$$
{\tt L}(\theta|\mX,\mY)=p(\mY|\mX,\theta,\sigma^2) = \phi_n(\mY; \mX\theta, \sigma^2 \mI_n)
= \frac{1}{\sqrt{(2\pi\sigma^2)^n}}\exp\left(-\frac{1}{2\sigma^2}(\mY-\mX\theta)^\top(\mY-\mX\theta)\right).
$$
The maximum likelihood estimator $\widehat{\theta}^{(\tt MLE)}$ of $\theta$ is the well-known
\begin{eqnarray}
\label{eq:MLE:theta:1}
\widehat{\theta}^{(\tt MLE)} = \underset{\theta \in \Theta}{\tt argmax}\left\{{\tt L}(\theta|\mX,\mY)\right\}=
(\mX^\top\mX)^{-1}\mX^\top\mY.
\end{eqnarray}
Combining the fact that $\widehat{\theta}^{(\tt MLE)}  \sim N_p(\theta, \sigma^2 (\mX^\top\mX)^{-1})$ with the conjugate prior $\theta \sim N_p(\theta_0, \sigma^2 \Lambda_0^{-1})$, the Bayesian estimator $\widehat{\theta}^{(\tt Bayes)}$ of the vector $\theta$ of regression coefficients, is given by
\begin{eqnarray}
\label{eq:Bayes:theta:1}
\widehat{\theta}^{(\tt Bayes)} = \mathbb{E}[\theta|\mY] = (\mX^\top\mX+\Lambda_0)^{-1}(\mX^\top\mX \widehat{\theta}^{(\tt MLE)} + \Lambda_0\theta_0).
\end{eqnarray}
The famous ridge regression estimator $\widehat{\theta}_{\lambda}^{(\tt ridge)}$ of $\theta$ first proposed by \cite{Ridge:Regression:Hoerl:1} and \cite{Ridge:Regression:Hoerl:2} is shown to be special case of the above Bayesian estimator when $\Lambda_0 = \lambda \mI$. Specifically,
\begin{eqnarray}
\label{eq:Ridge:theta:1}
\widehat{\theta}_{\lambda}^{(\tt ridge)} &=& \underset{\theta \in \Theta}{\tt argmin}\left\{(\mY-\mX\theta)^\top(\mY-\mX\theta) + \lambda\theta^\top\theta \right\} \nonumber \\
&=&(\mX^\top\mX+\lambda\mI)^{-1}\mX^\top\mY \nonumber \\
&=& \mathbb{E}[\theta|\mY] = \widehat{\theta}^{(\tt Bayes)}
\end{eqnarray}
It is easy to verify (check) that {\it the maximum likelihood estimator is a special case of the Bayesian estimator}, in the sense that
$$
\underset{\lambda \rightarrow \infty}{\lim} \widehat{\theta}_{\lambda}^{(\tt ridge)} =
\underset{\lambda \rightarrow \infty}{\lim} (\mX^\top\mX+\lambda\mI)^{-1}\mX^\top\mY =
(\mX^\top\mX)^{-1}\mX^\top\mY =\widehat{\theta}^{(\tt MLE)}.
$$
It also easy to see that the ridge estimator is a shrinkage estimator, with the tendency to shrink all
the components of the vector to zero together as $\lambda$ gets ever larger. Specifically,
$$
\underset{\lambda \rightarrow \infty}{\lim} \widehat{\theta}_{\lambda}^{(\tt ridge)} = {\bf 0}.
$$
Of great importance to big data analytics is the fact that between the two extremes of zero $\lambda$ and infinite
$\lambda$, lies a value of $\lambda$ that achieves the trade-off between bias and variance, and thereby achieves the smallest cross validation error.
%
The fact that the Bayesian estimator is inherently (by its very design) biased, used to be a subject of great debates, until numerous findings revealed that unbiasedness while a desired property, is not the be all and end all of statistical estimation and inference, quite far from it. It turns out that most scientific endeavors reveal the fundamental need for a trade-off between bias and variance. For the regression example mentioned earlier, when
we (a) either have multicollinearity in the design matrix or (b) the data matrix $\mX$ is high dimensional but with a very low sample size ($n \lll p$, underdetermined system), the maximum likelihood estimator is theoretical unbiased
but has an ill-conditioned variance matrix that leads to non-existence or non uniqueness or severe instability. Even in case where a numerical solution can realized, the variance is inflated because of the near singularity. The Bayesian approach via ridge regression for instance yields a solution, albeit biased, but with a reduced variance. In fact, in the $n \lll p$ it is impossible to have any solution without a device like the ridge approach. This is the kind of scenarios that make us say that the Bayesian paradigm is a gospel, meaning good news, as it allows workable solution where none appears to exist. Solutions like ridge are nowadays ubiquitous in statistical machine learning and belong to a class of machine learning approaches known as regularization methods, where all the techniques consist of adding constraints to an ill-posed problem to hopefully achieve well-posedness in Hadamard's sense. All the methods in the regularization framework are centered around the regularized empirical risk
$$
R_{\tt reg}(f) = \frac{1}{n}\sum_{i=1}^n{\ell(Y_i, f(X_i))} + \lambda \|f\|_{\mathcal{H}}
$$
where $\|f\|_{\mathcal{H}}$ is the norm of $f$ in the function space $\mathcal{H}$. For the linear regression learning task for instance, the ridge regularization mentioned earlier has evolved (been developed) alongside the Least Absolute Shrinkage and Selection Operator (LASSO) proposed by \cite{Tibshirani:1996} \cite{Tibshirani94regressionshrinkage}, which admits a Bayesian formulation using a Laplace prior on $\theta$, but does not yield a closed-form solution like the ridge estimator.
\begin{eqnarray}
\label{eq:lasso:theta:1}
\widehat{\theta}_{\lambda}^{(\tt lasso)} = \underset{\theta \in \Theta}{\tt argmin}\left\{(\mY-\mX\theta)^\top(\mY-\mX\theta) + \lambda\|\theta\|_1 \right\}
\end{eqnarray}

The well known greatest strength of the LASSO estimator comes from the fact that it does achieve sparsity and therefore is used for variable selection. Just like the
ridge solution, the LASSO, through regularization, is inherently able to yield a solution where the MLE would at best be very unstable.
\begin{figure}[!h]
  \centering
\includegraphics[width=7.5cm, height=6.5cm]{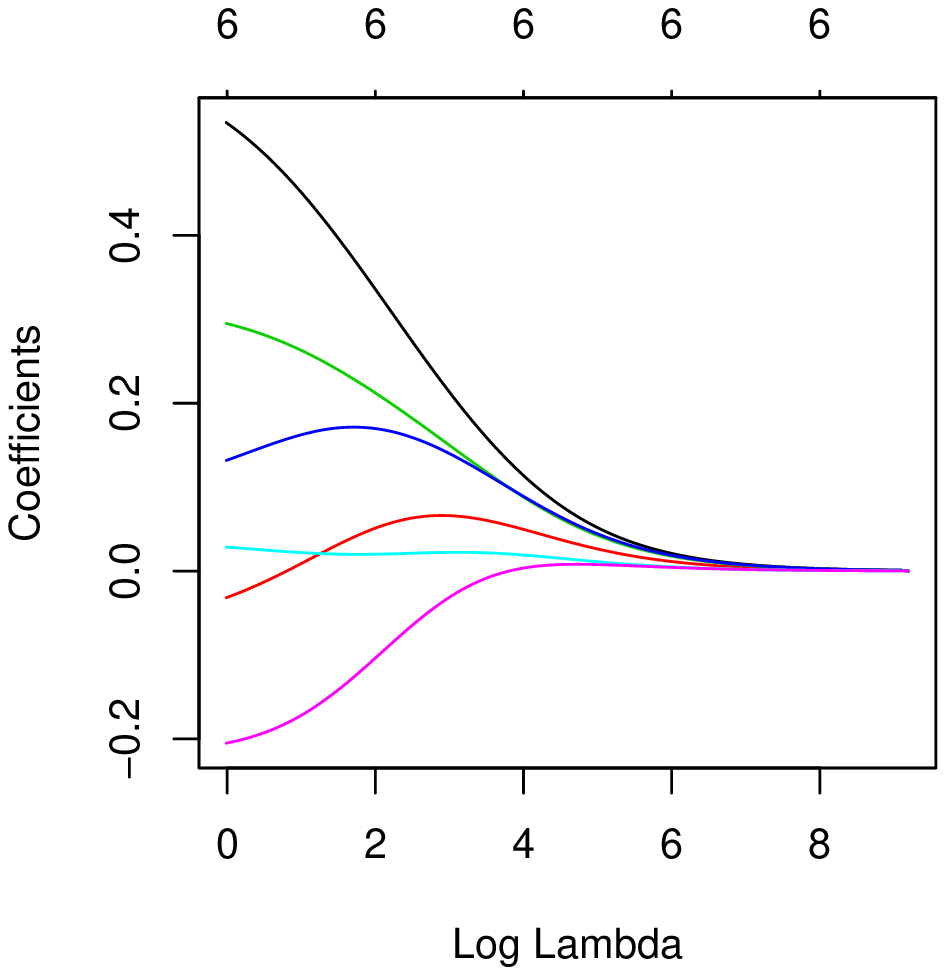}
\includegraphics[width=7.5cm, height=6.5cm]{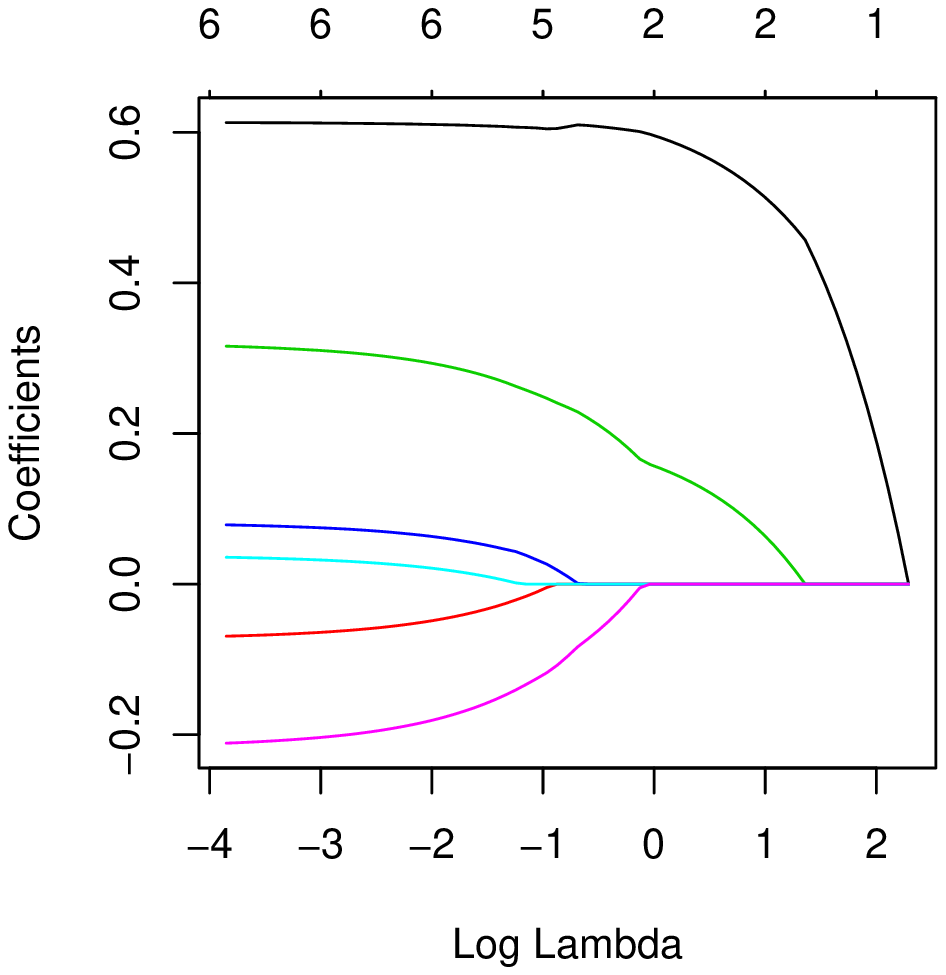}
  \caption{Depiction of the solution path: (left) ridge shrinkage  (right) lasso selection on the attitude dataset}
  \label{fig:ridge:lasso:shrinkage:1}
\end{figure}
It is interesting to see that the LASSO estimator does indeed select along with shrinkage whereas the ridge estimator
simply shrinks while maintaining all the six initial variables. Combining ridge and lasso, one gets
$$
R_{\tt reg}(f) = \frac{1}{n}\sum_{i=1}^n{\ell(Y_i, f(X_i))} + \lambda {\tt Penalty}_\alpha(\theta),
$$
where ${\tt Penalty}_\alpha(\theta) = (1-\alpha)\|\theta\|_1 + \alpha\|\theta\|_2$ is the so-called elastic net penalty. Several implementations exist in R, including \cite{Friedman:2010:1} and \cite{Friedman:2010:2}.
\begin{figure}[!h]
  \centering
\includegraphics[width=7.5cm, height=6.5cm]{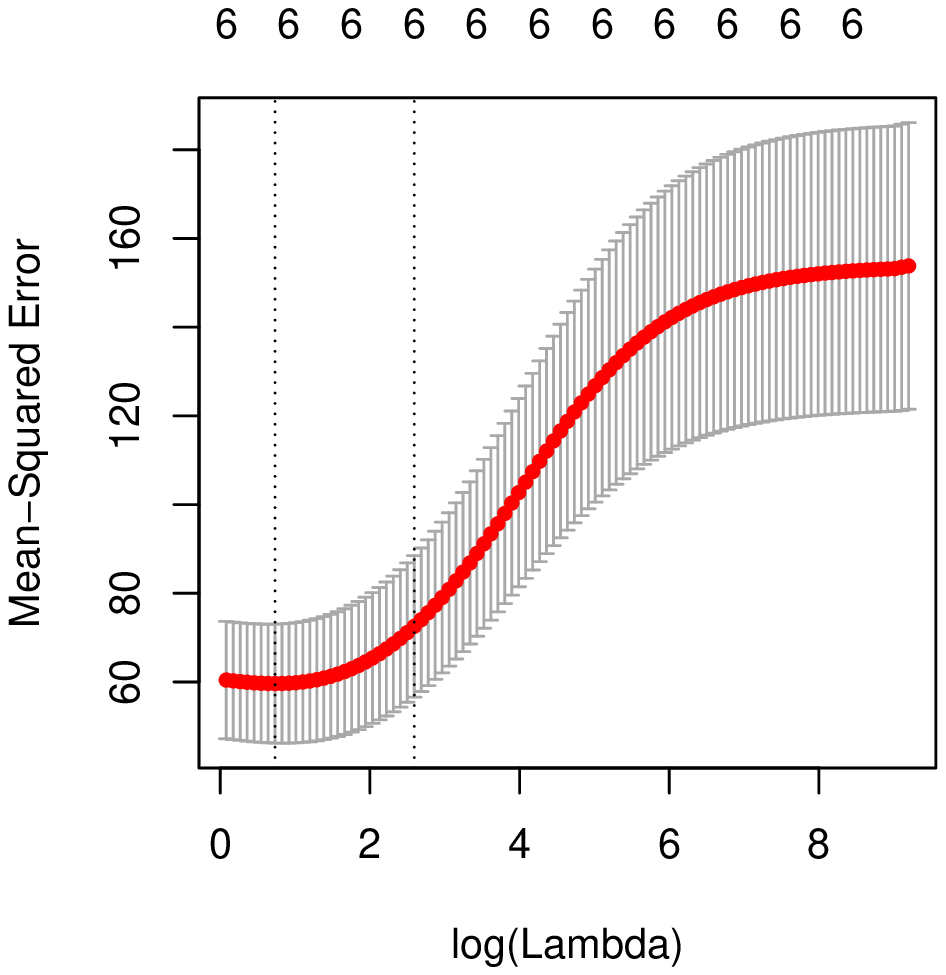}
\includegraphics[width=7.5cm, height=6.5cm]{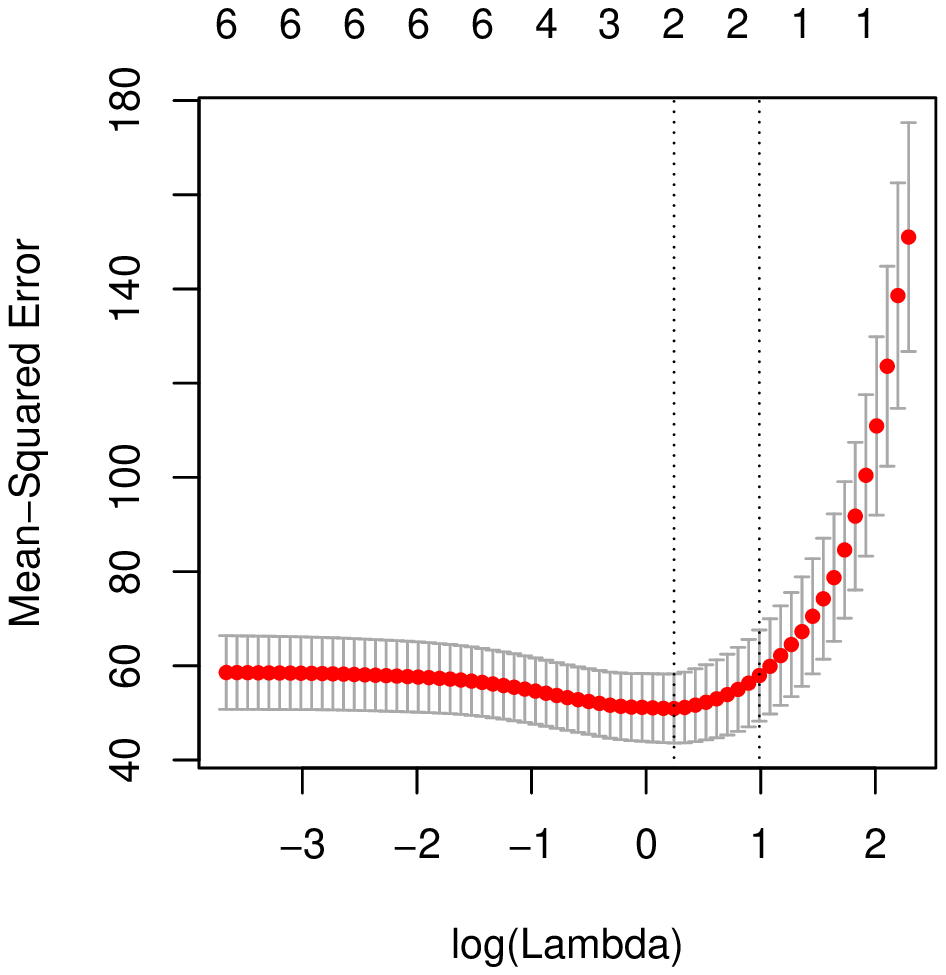}
  \caption{Depiction of the tuning of $\lambda$: (left) ridge  (right) lasso,  on the attitude dataset}
  \label{fig:ridge:lasso:cv:1}
\end{figure}
Figures \eqref{fig:ridge:lasso:shrinkage:1} and  \eqref{fig:ridge:lasso:cv:1} depict the application of the elastic net penalty to regression analysis learning task on a $6$-dimensional dataset known as the {\sf attitude}, where the goal is to regress the
{\it rating} of companies to six variables, namely {\sf complaints}, {\sf learning}, {\sf privileges}, {\sf raises}, {\sf critical}, {\sf advance}. Different prior probability distributions (ridge or lasso) on the parameter space, yield difference solution paths,
Figure \eqref{fig:ridge:lasso:shrinkage:1}. LASSO selects a small subset of variables, whereas ridge keeps them all, albeit while shrinking all of them together towards zero.
Once again, the goal of this paper, is not to explicate the technical niceties of the Bayesian paradigm, but instead draw the reader's attention on its immense modelling potential. As a matter of fact, the above regularization framework based on the elastic net, helps tackle and solve many predictive analytics  task in high dimension and low sample size situations as arises with DNA Gene Expression Microarray Data and several other large $p$ small $n$ tasks.
It bears repeating that all the above pearls of statistical modeling, while set up in the so-called regularization framework, do inherently admit a Bayesian formulation. Like we said earlier, "To Bayes or Not Bayes?" is no longer the question, but rather "How am I making the most of Bayes?". The key seems to lie in the specification of carefully thought out prior densities that allow one to isolate precisely the kind of solution desired out of the multiplicity of solutions.
{\it Wherever there is statistical learning, especially in settings where there is an ill-posedness challenge, the Bayesian paradigm is forever available as a formidable weapon in the statistical scientist's modelling arsenal.}
We see the power of the Bayesian thought directly or indirectly in state of the art settings such Latent Dirichlet Allocation (LDA) for Topic Modelling \cite{Blei:LDA:2003}. The forever useful Kalman filter, thanks to the latent space has also benefitted heavily from the power of the Bayesian paradigm \cite{Bayesian:Kalman:Filter:Wiley:1}, \cite{Bayesian:Kalman:Filter:Wiley:2} and \cite{Bayesian:Kalman:Filter:Wiley:3}.
Even before the blessings of affordable computation ushered in the glorious era of the Bayesian thought, Markov Random Fields were being used for the Statistical Analysis of Dirty Pictures \cite{Besag:1986}, already anchoring the palpable power of the Gospel according to Reverend Thomas Bayes. And yes, modern artificial intelligence as also benefitted very immensely from the flexibility that the combination of likelihood with prior affords the statistical scientists, in \cite{Bayesian:NeuralNets:Neal:1996}, we see that the explosion of Neural Networks as tools for artificial intelligence and learning was quickly found to have a nice connection to the Bayesian paradigm, and even now works like \cite{Bayesian:RecurrentNeuralNets:Fortunato:2017} demonstrate the great appeal of the Bayesian approach for the now very fashionable and in vogue  Networks as well. \cite{MacKay:1991:1} gives a detailed account of Bayesian interpolation and introduces the now popular and widely used concept of automatic relevance determination (ARD).
As a matter of fact, \cite{Tipping:01:1}'s Sparse Bayesian Learning and the Relevance Vector Machine (RVM) is a nice
piece of work inspired by a combination of \cite{MacKay:1991:1} and \cite{Vapnik:SVM:96}.
Interestingly, a little after \cite{Tipping:01:1}, we get \cite{Sollich2002BayesianMF} exploring Bayesian Methods for Support Vector Machines and more recently \cite{NIPS2014_5507} with an interesting account of Bayesian Nonlinear Support Vector Machine. \cite{Williams96gaussianprocesses}'s work on Gaussian process regression and later  
\cite{NIPS1999_1694} with Efficient Algorithms for Bayesian Gaussian Processes, both ushered in a series of contributions in machine learning featuring Bayesian Gaussian processes for regression and classification, later
crystalized in \cite{Rasmussen:Williams:2006} which has become one of the main textbook for the use of Bayesian Gaussian processes in machine learning. 
\cite{Csato:Fokoue:2000} explore ideas of variational mean field approximations featuring efficient Approaches to Bayesian Gaussian Process Classification. 
Gaussian Process Priors open the door to a vast universe of nonparametric statistical modelling in the Bayesian framework.
This use of prior distributions over function spaces central to Gaussian process learning has recently become mainstream in Bayesian Nonparametric statistical analysis, anchored by the seminal
work on the introduction of the Dirirchlet process prior by \cite{Ferguson:Dirichlet:Process:1973} and \cite{Ferguson:Dirichlet:Process:1974}
which has enriched the statistician and data scientist's modelling arsenal with a formidably powerful weapon in nonparametric statistical analysis, especially allowing  prior distributions on function spaces and infinite dimensional spaces in general. The recent years have been marked by what is literally an explosion of methods which 
are derivatives of or inspired by the seminal work of \cite{Ferguson:Dirichlet:Process:1973} and \cite{Ferguson:Dirichlet:Process:1974}. With Bayesian nonparametrics providing extra modelling strength and flexibility to statisticians and data scientists, the vast territory of application of the Bayesian paradigm just keeps on expanding, further justifying  our view that the Gospel according to Reverend Thomas Bayes is indeed ubiquitous,
pervading and permeating the whole of science. Statistical Machine Learning from its very early days both implicit and explicitly gave  a prominent platform and a loud speaking voice to the Bayesian school of thought, gaussian processes and Dirichlet processes have increased the volume of the loud speaker.
In Bayesian computation, the 1990 seminal work  of \cite{Gelfand:Smith:Gibbs:1990} introduce the world to the power
of the Gibbs Sampler, and made  it possible for Bayesian statisticians to tackle and successfully solve many statistical modeling problems which had eclipsed them until that milestone.
After the \cite{Gelfand:Smith:Gibbs:1990} paper that launched the Bayesian Computation revolution, software packages like {\bf BUGS}({\tt Bayesian Inference with the Gibbs Sampler}) began to emerge, making it more and more possible
for Bayesians to actually solve interesting and meaningful real life problems. Implementations abound that help practitioners experiment and applied the power of the Gibbs sampler \cite{Plummer:JAGS:Gibbs:2003}. The statistical software environment {\sf R} has many packages and an entire view {\sf install.view(Bayesian)}
that contain various functions for Bayesian analyses of all kinds. As a matter of fact, with the development of
Bayesian computation which marked the birth of a collection of methods known as Markov Chain Monte Carlo (MCMC) methods, literally every aspect of statistical benefitted from the modelling power of the Bayesian paradigm.
The development of Bayesian computation also allowed substantial progress in Bayesian model selection and Bayesian
variable selection. Among other contributions, we have Spike and Slab \cite{Ishwaran:SpikeSlab:2005} and more
recently  work on featuring mixtures of g-Priors for Bayesian variable selection  \cite{Liang08mixturesof}. 
The Bayesian Model Averaging for Linear Regression Models by \cite{Raftery:BMA:1997} was later supplemented by
a tutorial \cite{Hoeting:BMA:1999}, that further helped put practical BMA on a firm foundation. Later, 
\cite{Barbieri02optimalpredictive} provided optimal predictive model selection via the so-called 
Bayesian median model. The Estimation of Atom Prevalence for Optimal Prediction \cite{Fokoue:2008:2}
sought to be a flexible and more adaptive counterpart to  \cite{Barbieri02optimalpredictive}
As we said earlier, the intention of this paper is far from any attempt to provide an exhaustive technical exploration of the Bayesian paradigm. That would be gargantuan and virtually impossible. Instead, we have sought throughout and hope to have given the reader a visceral sense of the appeal of the Bayesian paradigm as a statistical machine learning tool for data science. We complete by mentioning a few contributions of the Bayesian paradigm to latent variable modelling and kernel regression, with works like 
\cite{Fokoue:2011:1} which introduces a stable Radial Basis Function Selection via Mixture Modelling of the Sample Path,  and \cite{Fokoue:Sun:Goel:2011:1} that extends it with a fully Bayesian Analysis of the Relevance Vector Machine With Extended Prior. \cite{Fokoue:2009:1} proposes and develops a Bayesian computation of the Intrinsic Structure of Factor Analytic Models, drawing some of its elements from 
\cite{Fokoue:Titterington:2003:1} where mixtures of Factor Analysers featuring Bayesian Estimation and Inference by Stochastic Simulation.
 
\section{Conclusion and discussion}
In this paper, we have provided a general bird's eye view of the manifold ways in which the Bayesian
paradigm has become one of the main tools in the arsenal of all statisticians and data scientists.
In our experience and observation, human statistical thought and perception as witnessed in interval estimation and hypothesis testing appears to be inherently and quintessentially Bayesian: indeed when non statisticians are asked to interpret confidence intervals or pvalues, most (if not all) say things that are essentially credible sets or posterior probabilities of hypotheses. It does seem that our human statistical thought quintessentially agrees with the Bayesian principle.
When it comes to decision making under uncertainty,
It is virtually impossible to find a field of study in science or otherwise that has not been heavily and positively by the power of the Bayesian paradigm.
Uncertainty is ideally dealt with using the powerful language of probability, hence the appropriateness of the assignment of property to all unknown quantities, including parameters unfortunately treated by others as fixed.
The Bayesian approach is the only way to properly deal with latent variable models. What other way exists to properly model a random variable other than specify or estimate its distribution from the data
Penalized Least Squares Estimation turns out to admit a natural Bayesian formulation with penalties capturing a prior belief about distributional aspects of the parameters or function class of interest. In sensu lato and senso stricto, the likelihood principle is a special case (subset) of the Bayesian paradigm.
The inherent capacity of the Bayesian paradigm to extend the likelihood principle can be likened to the way in which the relativity theory contributed by Albert Einstein extended, enriched and revolutionized Isaac Newton's fundamental laws of physics.
The gospel according to Reverend Thomas Bayes lives on and keeps on gaining more power and transform more lives through its impact in science, statistical machine learning and data science
From an unpublished manuscript, the revolutionary idea of Reverend Thomas Bayes has become one of the most consequential and most pervading transformative concepts in the whole of science and epistemology. Indeed, 
wherever there is a bona fide likelihood, there is room for Bayes.

\bibliographystyle{chicago}
\bibliography{bayesian-ubiquitous-1}

\end{document}